\documentclass[twocolumn,showpacs,showkeys,preprintnumbers,amssymb,aps,superscriptaddress,pre]{revtex4}
\usepackage{graphicx}
\usepackage{dcolumn}
\usepackage{bm}

\usepackage{epsfig}
\usepackage{color}

\begin{document}

\preprint{APS/123-QED}

\title{Magnetization process, bipartite entanglement and enhanced magnetocaloric effect of the exactly solved spin-1/2 Ising-Heisenberg tetrahedral chain}

\author{Jozef Stre\v{c}ka}
\email{jozef.strecka@upjs.sk}
\affiliation{Department of Theoretical Physics and Astrophysics, Faculty of Science, 
P. J. \v{S}af\'{a}rik University, Park Angelinum 9, 040 01, Ko\v{s}ice, Slovak republic}
\author{Onofre Rojas}
\affiliation{Departamento de Ciencias Exatas, Universidade Federal de Lavras, 37200-000, Lavras-MG, Brazil}
\author{Taras Verkholyak}
\affiliation{Institute for Condensed Matter Physics, National Academy of Sciences of Ukraine, 1 Svientsitskii Street, L'viv-11, 79011, Ukraine}
\author{Marcelo L. Lyra}
\affiliation{Instituto de F\'isica, Universidade Federal de Alagoas, 57072-970, Maceio-AL, Brazil}

\date{\today}

\begin{abstract}
The frustrated spin-1/2 Ising-Heisenberg ladder with Heisenberg intra-rung and Ising inter-rung interactions is exactly solved in a longitudinal magnetic field by taking advantage of the local conservation of the total spin on each rung and the transfer-matrix method. We have rigorously calculated the ground-state phase diagram, magnetization process, magnetocaloric effect and basic thermodynamic quantities for the model, which can be alternatively viewed as an Ising-Heisenberg tetrahedral chain. It is demonstrated that a stepwise magnetization curve with an intermediate plateau at a half of the saturation magnetization is also reflected in respective stepwise changes of the concurrence serving as a measure of bipartite entanglement. The ground-state phase diagram and zero-temperature magnetization curves of the Ising-Heisenberg tetrahedral chain are contrasted with the analogous results of the purely quantum Heisenberg tetrahedral chain, which have been obtained through density-matrix renormalization group (DMRG) calculations. While both ground-state phase diagrams fully coincide in the regime of weak inter-rung interaction, the purely quantum Heisenberg tetrahedral chain develops Luttinger spin-liquid and Haldane phases for strongly coupled rungs which are absent in the Ising-Heisenberg counterpart model.
\end{abstract}


\maketitle

\section{Introduction}

Exactly solved quantum spin models represent important milestones in statistical mechanics as they provide deeper insights into otherwise hardly understandable aspects of cooperative phenomena \cite{matt93,lieb04,miya11}. The spin-$\frac{1}{2}$ quantum Ising and XY chains in a transverse magnetic field are for instance paradigmatic examples of exactly tractable models, which have helped us to clarify generic features of quantum phase transitions \cite{lieb61,kats62}. It should be emphasized, however, that the complete rigorous solution is beyond the scope of present knowledge for most of the quantum spin models.  

An important utilization of exactly solved quantum spin models has been recently found in the field of quantum information and computation, where they provide indispensable ground for the development and further testing of entanglement measures (see Refs. \cite{plen07,amic08,horo09} for comprehensive reviews on this subject). It is noteworthy that the quantum entanglement is the main resource that allows the quantum computation and communication \cite{niel00}, whereas entanglement measures can be related via certain witnesses to thermodynamic quantities that offer an intriguing possibility for experimental testing \cite{souz08,soar09,chak12}. Among the most challenging tasks in this research field is to explore a bipartite entanglement near quantum critical points of exactly solved spin chains \cite{osbo02,oste02,vida03,somm04,cao06,mazi10,roja12,werl13}. 

A significant variation of entanglement measures has been also found across the quantum phase transition of the antiferromagnetic spin-$\frac{1}{2}$ Heisenberg two-leg ladder driven by a longitudinal magnetic field \cite{trib09,bose05}, which exhibits a peculiar gapless Luttinger liquid phase at moderate magnetic fields. Beforehand, the quantum spin-$\frac{1}{2}$ Heisenberg two-leg ladder has been subject of tremendous theoretical efforts mainly due to a rather extensive list of solid-state materials (falling mostly into the family of cuprates) offering its experimental realizations \cite{dago96,dago99,batc07}. Later, it has been demonstrated that the frustrated spin-$\frac{1}{2}$ Heisenberg two-leg ladder with an additional diagonal (next-nearest-neighbor) coupling between two legs exhibits an intriguing intermediate plateau at half of the saturation magnetization and macroscopic magnetization jumps related with its presence \cite{mila98,hone00,chan06,mich10}. Bose and Chattopadhyay have proved that the macroscopic magnetization plateaux and jumps give rise to similar features of the entanglement measures \cite{bose02}.

A few rigorous results have been reported so far only for the special case of the frustrated spin-$\frac{1}{2}$ Heisenberg two-leg ladder with equal interactions along legs and diagonals, which has two different gapped rung singlet-dimer and Haldane-like zero-field ground states \cite{gelf91,xian94,xian95,suth00,ekim08}. It is worthwhile to remark that this notable special case can be alternatively viewed as the spin-$\frac{1}{2}$ Heisenberg chain of linked tetrahedra (tetrahedral chain). Recently, we have proposed the hybrid spin-$\frac{1}{2}$ Ising-Heisenberg two-leg ladder with the Heisenberg intra-rung and Ising inter-rung interactions, which is fully exactly tractable at zero \cite{verk12,verk13} as well as non-zero \cite{roja13} temperatures. The main goal of the present work is to clarify the magnetic behavior of this exactly solvable model in a longitudinal magnetic field, which may also reveal some important aspects of its full quantum Heisenberg counterpart. A possible experimental realization of the anisotropic version of the spin-$\frac{1}{2}$ Heisenberg tetrahedral chain has been recently discovered in Cu$_3$Mo$_2$O$_9$ \cite{hase08,kuro11,mats12}.

The organization of this paper is as follows. The Ising-Heisenberg tetrahedral chain is introduced in Sec. \ref{sec:ihm} along with basic steps of its exact analytical treatment. Sec. \ref{sec:hm} briefly describes the details of DMRG calculations performed for the analogous but purely quantum Heisenberg tetrahedral chain. The most interesting results for the ground state, magnetization process, bipartite entanglement and overall thermodynamics of the Ising-Heisenberg tetrahedral chain are discussed in Sec. \ref{sec:result}. Sec. \ref{sec:comp} further compares ground-state phase diagrams and magnetization process of the Ising-Heisenberg tetrahedral chain and its full quantum Heisenberg counterpart. Finally, several concluding remarks are mentioned in Sec. \ref{sec:conc}. 

\section{Ising-Heisenberg tetrahedral chain}
\label{sec:ihm}

\begin{figure}
\begin{center}
\includegraphics[width=0.45\textwidth]{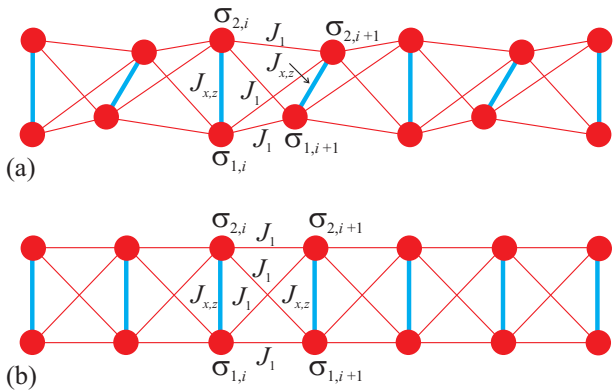}
\end{center}
\vspace{-0.6cm}
\caption{(a) A diagrammatic representation of the spin-$\frac{1}{2}$ Ising-Heisenberg tetrahedral chain. Thick (blue) lines represent the XXZ Heisenberg coupling ($J_x$, $J_z$), while thin (red) lines correspond to the Ising coupling $J_1$; (b) the tetrahedral chain can also be viewed as a frustrated two-leg ladder with equal Ising interactions along legs and diagonals.}
\label{fig1}
\end{figure}

Let us consider the spin-$\frac{1}{2}$ Ising-Heisenberg chain, which consists of edge-sharing tetrahedra involving Heisenberg and Ising interactions as schematically illustrated in Fig.~\ref{fig1}(a). It is quite apparent from this figure that the Heisenberg coupling is assigned to all common edges shared by neighboring tetrahedra, while all other edges schematically represent the Ising interaction. The Ising-Heisenberg tetrahedral chain can be alternatively viewed as a special case of the frustrated Ising-Heisenberg two-leg ladder \cite{verk12,verk13} with the Heisenberg intra-rung interaction and uniform Ising coupling along legs and diagonals, respectively (see Fig.~\ref{fig1}(b)). The Hamiltonian of the spin-$\frac{1}{2}$ Ising-Heisenberg tetrahedral chain is given by
\begin{eqnarray}
\label{eq:Ham-orig}
\hat{\cal H} \!\!&=&\!\! \sum_{i=1}^{N}\left[ J_x (\hat{\sigma}_{1,i}^{x}\hat{\sigma}_{2,i}^{x} + \hat{\sigma}_{1,i}^{y}\hat{\sigma}_{2,i}^{y}) 
                    + J_{z}\hat{\sigma}_{1,i}^{z}\hat{\sigma}_{2,i}^{z}\right. \\
 \!\!&+&\!\! \left.J_{1}(\hat{\sigma}_{1,i}^{z}+\hat{\sigma}_{2,i}^{z})(\hat{\sigma}_{1,i+1}^{z}+\hat{\sigma}_{2,i+1}^{z}) 
        - h (\hat{\sigma}_{1,i}^{z}+\hat{\sigma}_{2,i}^{z}) \right],
\nonumber
\end{eqnarray}
where $\hat{\sigma}_{\gamma,i}^{\alpha}$ marks the standard spin-$\frac{1}{2}$ Pauli operator, the superscript $\alpha \in \{x,y,z\}$ labels its spatial component and the subscript $\gamma = 1,2$ specifies a given leg (see Fig. \ref{fig1}(b)). The interaction terms $J_x$ and $J_{z}$ denote the spatially anisotropic XXZ Heisenberg interaction incident to the edges shared by the neighboring tetrahedra, the parameter $J_{1}$ labels the Ising interaction attached to all the other (not shared) edges and $h$ is Zeeman's term associated with a presence of the longitudinal magnetic field. It is worthwhile to remark that different sign convention is used in the Hamiltonian (\ref{eq:Ham-orig}) for coupling constants as compared to our preceding paper concerned with the zero-field limit of the model under investigation \cite{roja13}, since our attention will be henceforth restricted only to a particular case of the model with the antiferromagnetic interactions $J_x>0$, $J_z>0$ and $J_1>0$ displaying the most interesting physical features such as the spin frustration, magnetization plateaux, etc.

As it has been shown previously \cite{roja13}, it is quite convenient to rewrite the Hamiltonian (\ref{eq:Ham-orig}) with the help of spatial components of the total spin angular momentum $\hat{S}_{i}^{\alpha}=\hat{\sigma}_{1,i}^{\alpha}+\hat{\sigma}_{2,i}^{\alpha}$ of two spins coupled by the Heisenberg interaction. Using the spin identity $(\hat{S}_{i}^{\alpha})^{2}=\frac{1}{2}+2\hat{\sigma}_{1,i}^{\alpha}\hat{\sigma}_{2,i}^{\alpha}$, the Hamiltonian (\ref{eq:Ham-orig}) of the spin-$\frac{1}{2}$ Ising-Heisenberg tetrahedral chain takes the form  
\begin{eqnarray}
\hat{\cal H} \!\!&=&\!\! \sum_{i=1}^{N} \left[ J_{1} \hat{S}_{i}^{z} \hat{S}_{i+1}^{z} + \frac{J_x}{2} \boldsymbol{\hat{S}}_{i}^{2} + \frac{J_{z}-J_x}{2} \left(\hat{S}_{i}^{z}\right)^{2} - h \hat{S}_i^z \right] \nonumber \\
\!\!&-&\!\! \frac{N}{4}(2J_x+J_{z}).
\label{eq:Ham-eff}
\end{eqnarray}
Apparently, the Hamiltonian (\ref{eq:Ham-eff}) is entirely expressed in terms of the total spin angular momentum $\boldsymbol{\hat{S}}_{i}^{2}$ and its $z$-spatial component $\hat{S}_{i}^{z}$, which correspond to conserved quantities with well defined quantum numbers $S_i (S_i + 1)$ and $S_i^z = -S_i, -S_i + 1, \ldots, S_i$, respectively. Hence, it follows that the Hamiltonian (\ref{eq:Ham-orig}) of the spin-$\frac{1}{2}$ Ising-Heisenberg tetrahedral chain has been rigorously mapped onto the Hamiltonian (\ref{eq:Ham-eff}) of some classical composite spin-chain model as the quantum spin number for the total angular momentum of two spins is either $S_i = 0$ or $1$. It is quite obvious from Eq. (\ref{eq:Ham-eff}) that the Ising interaction $J_1$ and the external magnetic field $h$ directly determine the effective nearest-neighbor interaction and effective field within the composite spin-chain model. In addition, the XXZ Heisenberg interaction determines an effective single-ion anisotropy $\frac{J_{z}-J_x}{4}$ acting within the triplet sector of the composite spin $S_i = 1$ and it also shifts the energy between singlet and triplet sectors. Bearing all this in mind, the Hamiltonian of the spin-$\frac{1}{2}$ Ising-Heisenberg tetrahedral chain can be fully diagonalized and rewritten into the form 
\begin{eqnarray}
{\cal H} = {\cal H}_0 + \sum_{i=1}^{N} {\cal H}_i, 
\label{eq:diagham}
\end{eqnarray}
which is split into the less important constant term ${\cal H}_0 = -\frac{N}{4}(2J_x+J_{z})$ and the sum of symmetric expressions ${\cal H}_i$ involving only eigenvalues of two subsequent composite spins $S_{i}$ and $S_{i+1}$ 
\begin{eqnarray}
{\cal H}_i \!\!&=&\!\! J_{1} S_{i}^{z}S_{i+1}^{z} + \frac{J_x}{4} \left[S_i(S_i + 1) + S_{i+1}(S_{i+1} + 1) \right] \nonumber   \\
\!\!&+&\!\! \frac{J_{z}-J_x}{4}\left[\left(S_{i}^{z}\right)^{2}+\left(S_{i+1}^{z}\right)^{2}\right] - \frac{h}{2} \left(S_{i}^{z} + S_{i+1}^{z} \right)\!. 
\label{eq:diagham1}
\end{eqnarray}
It is noteworthy that the diagonalized Hamiltonian (\ref{eq:diagham1}) of the composite spin-chain model can be straightforwardly used to obtain all available ground states, the overall excitation spectrum and besides, it also implies a possibility for an implementation of the classical transfer-matrix method \cite{baxt82} to gain exact results for thermodynamic quantities. Substituting the effective Hamiltonian (\ref{eq:diagham}) into the definition of the partition function, one indeed obtains the following relation for the partition function of the spin-$\frac{1}{2}$ Ising-Heisenberg tetrahedral chain  
\begin{eqnarray}
\mathcal{Z} \!\!&=&\!\! \mathrm{Tr} \, \mathrm{e}^{-\beta \hat{\cal H}} = \mathrm{e}^{-\beta {\cal H}_0} \sum_{\{ S_i \}} \prod_{i=1}^{N} \mathrm{e}^{-\beta {\cal H}_i} \label{pf} \\
\!\!&=&\!\! \mathrm{e}^{-\beta {\cal H}_0} \sum_{\{ S_i \}} \prod_{i=1}^{N} \boldsymbol{T} (S_i, S_i^z; S_{i+1}, S_{i+1}^z) 
       = \mathrm{e}^{-\beta {\cal H}_0} \mathrm{Tr} \, \boldsymbol{T}^N,
\nonumber 
\end{eqnarray}
where $\beta=1/(k_{\rm B} T)$, $k_{\rm B}$ is the Boltzmann's constant, $T$ is the absolute temperature, the summation $\sum_{\{ S_i \}}$ runs over all possible values of a full set of the quantum spin numbers $\{ S_i \}$ and the expression $\boldsymbol{T} (S_i, S_i^z; S_{i+1}, S_{i+1}^z)$ marks the transfer matrix 
\begin{eqnarray}
\boldsymbol{T} \!\!\!\!\!&&\!\!\!\!\! (S_i, S_i^z ; S_{i+1}, S_{i+1}^z) = \langle S_i, S_i^z| \mathrm{e}^{-\beta {\cal H}_i} | S_{i+1}, S_{i+1}^z \rangle \nonumber \\
\!\!\!\!&=&\!\!\!\! \left(\begin{array}{cccc}
1 & xyw & {x}^{2} &  xyw^{-1} \\
xyw & {x}^{2}{y}^{2}zw^2 & {x}^{3}yw & {x}^{2}{y}^{2}z^{-1}\\
{x}^{2} & {x}^{3}yw & {x}^{4} &  {x}^{3}yw^{-1}\\
xyw^{-1} & {x}^{2}{y}^{2}z^{-1} & {x}^{3}yw^{-1} & {x}^{2}{y}^{2}zw^{-2}
\end{array}\right)
\label{tm}
\end{eqnarray}
with $x=\mathrm{e}^{-\frac{\beta J_x}{4}}$, $y=\mathrm{e}^{-\frac{\beta J_z}{4}}$, $z=\mathrm{e}^{-\beta J_{1}}$, and $w=\mathrm{e}^{\frac{\beta h}{2}}$. Now, it is sufficient to find the largest eigenvalue of the transfer matrix (\ref{tm}) in order to gain an exact expression for the partition function, free energy and overall thermodynamics. By inspection, one out of four eigenvalues of the transfer matrix (\ref{tm}) equals zero ($\lambda_0 = 0$) since its first and third row are linearly dependent, while other three eigenvalues can be found by solving the cubic equation
\begin{equation}
\lambda^3 - a \lambda^2 + b \lambda + c =0
\label{eq:cubic-eq}
\end{equation}
with the coefficients
\begin{eqnarray}
a \!\!&=&\!\! 1 + \mathrm{e}^{-\beta J_x} + 2 \mathrm{e}^{-\beta J_1 - \frac{\beta}{2} (J_x + J_z)} \cosh (\beta h), \nonumber \\
b \!\!&=&\!\! 2 (-1 + \mathrm{e}^{-\beta J_1})(1 + \mathrm{e}^{-\beta J_x}) \mathrm{e}^{-\frac{\beta}{2} (J_x + J_z)} \cosh (\beta h) \nonumber \\
  \!\!&-&\!\! 2 \mathrm{e}^{-\beta (J_x + J_z)} \sinh (2 \beta J_1), \nonumber \\
c \!\!&=&\!\! 2 (1 + \mathrm{e}^{-\beta J_x}) \mathrm{e}^{-\beta (J_x + J_z)} [\sinh (2 \beta J_1) - \sinh (\beta J_1)].
\label{eq:abc}
\end{eqnarray}
Following the standard procedure (see for instance Ref.~\cite{rekt69}) one easily finds the remaining three transfer-matrix eigenvalues as roots of the cubic equation (\ref{eq:cubic-eq})
\begin{eqnarray}
\lambda_{j} =  \frac{a}{3} + 2 \mbox{sgn}(q) \sqrt{p} \cos \left[\frac{1}{3} \left(\phi + j 2 \pi \right)\right]\!, \, (j=1,2,3)
\label{rce}
\end{eqnarray}
whereas 
\begin{eqnarray}
p \!\!&=&\!\! \left( \frac{a}{3} \right)^2 - \frac{b}{3}, \nonumber \\
q \!\!&=&\!\! \left( \frac{a}{3} \right)^3 - \frac{ab}{6} - \frac{c}{2}, \nonumber \\
\phi \!\!&=&\!\! \arctan \left(\frac{\sqrt{p^3 - q^2}}{q} \right).
\label{rcec}
\end{eqnarray}
In the thermodynamic limit $N \to \infty$, the free energy per unit cell is determined just by the largest eigenvalue 
$\lambda_{\rm max} = {\rm max} \{\lambda_1, \lambda_2, \lambda_3\}$ among the three roots (\ref{rce}) of the cubic equation (\ref{eq:cubic-eq})
\begin{equation}
f = - \beta^{-1} \lim_{N \to \infty} \frac{1}{N} \ln {\cal Z} = -\frac{2J_x+J_{z}}{4}- \beta^{-1} \ln \lambda_{\rm max}.
\label{fe} 
\end{equation}

Next, let us calculate a few quantities such as the magnetization and pair correlation functions, which will shed light on a magnetic behavior of the investigated spin-chain model. The single-site magnetization normalized with respect to its saturation value ($m_s = \frac{1}{2}$) can be obtained by differentiating of the free energy with respect to the external magnetic field, which is equivalent to a differentiation of logarithm of the largest transfer-matrix eigenvalue easily performed with the help of Eq. (\ref{eq:cubic-eq})  
\begin{equation}
\frac{m}{m_s} = - \frac{\partial f}{\partial h} = \frac{\partial \ln \lambda_{\rm max}}{\partial (\beta h)} 
              = \frac{a_h \lambda_{\rm max}- b_h}{3\lambda_{\rm max}^2-2a\lambda_{\rm max}+b}.
\label{mag} 
\end{equation}
Here, we have introduced the expressions $a_h$ and $b_h$ defined as follows
\begin{eqnarray}
a_h \!\!&=&\!\! 2 \mathrm{e}^{-\beta J_1 - \frac{\beta}{2} (J_x + J_z)} \sinh (\beta h), \nonumber \\
b_h \!\!&=&\!\! 2 (-1 + \mathrm{e}^{-\beta J_1})(1 + \mathrm{e}^{-\beta J_x}) 
                     \mathrm{e}^{-\frac{\beta}{2} (J_x + J_z)} \sinh (\beta h). \nonumber \\
\label{azbz}
\end{eqnarray}
A similar procedure can be also employed for the calculation of spatial components of the pair correlation function 
between two spins coupled by the Heisenberg interaction
\begin{eqnarray}
C_{\alpha} \equiv \langle \hat{\sigma}_{1,i}^{\alpha} \hat{\sigma}_{2,i}^{\alpha} \rangle = 
- \frac{1}{4} + \frac{a_{\alpha} \lambda_{\rm max}^2 - b_{\alpha} \lambda_{\rm max} - c_{\alpha}}{3 \lambda_{\rm max}^3 - 2 a \lambda_{\rm max}^2 + b \lambda_{\rm max}}, 
\label{cor} 
\end{eqnarray}
which are expressed in terms of the coefficients $a_{\alpha}$, $b_{\alpha}$ and $c_{\alpha}$ explicitly given for $\alpha = x$ and $z$ by  
\begin{eqnarray}
a_x \!\!&=&\!\! \frac{1}{2} \mathrm{e}^{-\beta J_x} + \frac{1}{2} \mathrm{e}^{-\beta J_1 - \frac{\beta}{2} (J_x + J_z)} \cosh (\beta h), \nonumber \\
b_x \!\!&=&\!\! \frac{1}{2} (-1 + \mathrm{e}^{-\beta J_1})(1 + \mathrm{e}^{-\beta J_x}) \mathrm{e}^{-\frac{\beta}{2} (J_x + J_z)} \cosh (\beta h) \nonumber \\
    \!\!&+&\!\! (-1 + \mathrm{e}^{-\beta J_1}) \mathrm{e}^{-\frac{\beta}{2} (3J_x + J_z)} \cosh (\beta h) \nonumber \\
    \!\!&-&\!\! \mathrm{e}^{-\beta (J_x + J_z)} \sinh (2 \beta J_1), \nonumber \\
c_x \!\!&=&\!\! (1 + 2 \mathrm{e}^{-\beta J_x}) \mathrm{e}^{-\beta (J_x + J_z)} [\sinh (2 \beta J_1) - \sinh (\beta J_1)], \nonumber \\
a_z \!\!&=&\!\! \mathrm{e}^{-\beta J_1 - \frac{\beta}{2} (J_x + J_z)} \cosh (\beta h), \nonumber \\
b_z \!\!&=&\!\! (-1 + \mathrm{e}^{-\beta J_1})(1 + \mathrm{e}^{-\beta J_x}) \mathrm{e}^{-\frac{\beta}{2} (J_x + J_z)} \cosh (\beta h) \nonumber \\
    \!\!&-&\!\! 2 \mathrm{e}^{-\beta (J_x + J_z)} \sinh (2 \beta J_1), \nonumber \\
c_z \!\!&=&\!\! 2 (1 + \mathrm{e}^{-\beta J_x}) \mathrm{e}^{-\beta (J_x + J_z)} [\sinh (2 \beta J_1) - \sinh (\beta J_1)]. \nonumber \\
\label{corab} 
\end{eqnarray}
After evaluating the single-site magnetization (\ref{mag}) and both spatial components of the pair correlation function (\ref{cor}), one may obtain in a rather straightforward manner a concurrence that serves as a measure of bipartite entanglement between the nearest-neighbor spins coupled by the XXZ Heisenberg interaction \cite{woot98,amic04,oste13} 
\begin{eqnarray}
{\cal C} = {\rm max} \left\{0 , 4 |C_x| - 2 \sqrt{\left( \frac{1}{4} + C_z \right)^2 - m^2} \right\}.
\label{conc} 
\end{eqnarray} 
In the following, we will use the explicit formula (\ref{conc}) for the concurrence in order to quantify thermal entanglement between two spins coupled by the Heisenberg interaction at finite temperatures and magnetic fields. Note furthermore that any other pair of spins is completely disentangled on behalf of a lack of quantum correlations in between them. Last but not least, other important thermodynamic quantities like for instance entropy or specific heat can also be readily computed from the exact expression (\ref{fe}) for the reduced free energy by making use of standard thermodynamic relations
\begin{eqnarray}
S = -\frac{\partial f}{\partial T} \quad \mbox{and} \quad C = T \frac{\partial S}{\partial T}.
\label{sc} 
\end{eqnarray}

\section{Heisenberg tetrahedral chain}
\label{sec:hm}

It might be quite interesting to contrast the exact results derived previously for the spin-$\frac{1}{2}$ Ising-Heisenberg tetrahedral chain with the accessible results for the analogous but fully quantum spin-$\frac{1}{2}$ Heisenberg tetrahedral chain, which is defined through the Hamiltonian 
\begin{eqnarray}
\hat{\cal H} = \sum_{i=1}^{N} [ J \hat{\boldsymbol \sigma}_{1,i} \cdot \hat{\boldsymbol \sigma}_{2,i} 
     \!\!&+&\!\! J_{1}(\hat{\boldsymbol \sigma}_{1,i} + \hat{\boldsymbol \sigma}_{2,i}) \cdot (\hat{\boldsymbol \sigma}_{1,i+1} + \hat{\boldsymbol \sigma}_{2,i+1})  \nonumber \\
     \!\!&-&\!\! h (\hat{\sigma}_{1,i}^{z}+\hat{\sigma}_{2,i}^{z})].
\label{hh}
\end{eqnarray}
The interaction terms $J$ and $J_1$ have similar meaning as described previously with exception that they both refer to the isotropic Heisenberg coupling. It is worthwhile to recall that the zero-field limit of the model (\ref{hh}) has been thoroughly investigated, sometimes referred to as Gelfand ladder \cite{gelf91,chan06}, and an exact dimerized (rung singlet) ground state has been verified in the frustrated regime $\frac{J_1}{J} < 0.7135$. {\color{red} It should be also mentioned that the ground-state phase diagram and magnetization process of the spin-$\frac{1}{2}$ Heisenberg tetrahedral chain given by the Hamiltonian (\ref{hh}) have been already obtained in Ref. \cite{hone00} by employing density-matrix renormalization group (DMRG) calculations for the effective Hamiltonian derived in terms of the composite spin operators \cite{lege97}. In the following, we will present an alternative approach how to arrive to the same effective Hamiltonian by making use of projection operators \cite{xian94}.}

To treat the frustrated spin-$\frac{1}{2}$ Heisenberg ladder (\ref{hh}) within the DMRG method, it is advisable to derive first the effective Hamiltonian in the basis spanned over eigenstates $|0\rangle_i=(|\!\!\uparrow_{1,i}\downarrow_{2,i}\rangle - |\!\!\downarrow_{1,i}\uparrow_{2,i}\rangle)/\sqrt{2}$, $|1\rangle_i=|\!\!\uparrow_{1,i}\uparrow_{2,i}\rangle$, $|2\rangle_i=(|\!\!\uparrow_{1,i}\downarrow_{2,i}\rangle + |\!\!\downarrow_{1,i}\uparrow_{2,i}\rangle)/\sqrt{2}$, $|3\rangle_i=|\!\!\downarrow_{1,i}\downarrow_{2,i}\rangle$ of the decoupled spin-$\frac{1}{2}$ Heisenberg dimers (i.e. the $J_1 = 0$ limit) \cite{xian94,xian95}. All spin operators entering the Hamiltonian (\ref{hh}) can be subsequently expressed in terms of operators $A_{kl}$ ($k,l=0,1,2,3$), which obey the pseudospin algebra $[A_{kl}, A_{mn}]=A_{kn}\delta_{lm}-A_{ml}\delta_{nk}$ and have physical meaning of projection operators within this basis set (for example, $A_{00}$ is the projection operator for the singlet-dimer state, see Ref. \cite{xian94} for more details). The Hamiltonian (\ref{hh}) can be then fully expressed in terms of the projection operators 
\begin{eqnarray}
\hat{\cal H} \!\!&=&\!\! \sum_{i=1}^{N} \Bigl\{ J \left(\frac{1}{4} - A_{00}^i \right) + J_{1} [(A_{11}^i-A_{33}^i)(A_{11}^{i+1}-A_{33}^{i+1})  \nonumber \\
\!\!&+&\!\! (A_{21}^i+A_{32}^i)(A_{12}^{i+1}+A_{23}^{i+1}) + (A_{12}^i+A_{23}^i) \nonumber \\
\!\!&\times&\!\! (A_{21}^{i+1}+A_{32}^{i+1})] - h (A_{11}^i-A_{33}^i) \Bigr\}.
\label{hhpo}
\end{eqnarray}
The part of the Hamiltonian (\ref{hhpo}) depending on the projection spin operators $A_{kl}$ that act only within the triplet sector ($k,l=1,2,3$) 
can be subsequently replaced with the new spin-1 operators
\begin{eqnarray}
\hat{P_i}^{+} \!\!&=&\!\! \sqrt{2} (A_{12}^i + A_{23}^i), \nonumber \\ 
\hat{P_i}^{-} \!\!&=&\!\! \sqrt{2} (A_{21}^i + A_{32}^i), \nonumber \\ 
\hat{P_i}^{z} \!\!&=&\!\! A_{11}^i - A_{33}^i,
\label{s1}
\end{eqnarray}
which satisfy the usual SU(2) algebra with the standard commutation relations $[\hat{P_i}^{+}, \hat{P_i}^{-}] = 2 \hat{P_i}^{z}$ and $[\hat{P_i}^{z}, \hat{P_i}^{\pm}] = \pm\hat{P_i}^{\pm}$. Omitting the unimportant constant term $\frac{N J}{4}$, the Hamiltonian (\ref{hhpo}) can be modified to the following final form by taking into account the definition (\ref{s1}) of the spin-1 operators
\begin{eqnarray}
\hat{\cal H} = - J \sum_{i=1}^{N} A_{00}^i + J_{1} \sum_{i=1}^{N} \hat{\mathbf P}_i \cdot \hat{\mathbf P}_{i+1} - h \sum_{i=1}^{N} \hat{P}_i^{z}.
\label{hef}
\end{eqnarray}
It should be emphasized that the Hamiltonian (\ref{hef}) consists of three commuting parts. The first part involves a sum of singlet projection operators $A_{00}^i$ for individual rungs, the second part corresponds to the spin-1 Heisenberg chain and the third part accounts for the effect of the external magnetic field. Note that $A_{00}^i$ and $\hat{\mathbf P}_i$ act in the mutually orthogonal subspaces corresponding to the singlet and triplet states on $i$th rung. According to this, we have to minimize the overall energy as a sum of three separate contributions considered above in order to get the ground-state energy of the fully quantum spin-$\frac{1}{2}$ Heisenberg tetrahedral chain. While the first contribution lowers the overall energy by $-J$ for each rung residing the singlet-dimer state, one may take advantage of the state-of-art DMRG calculations in order to get an accurate estimate of the ground-state energy of the effective spin-1 Heisenberg chain \cite{whit93}. This procedure allows us to avoid extensive (more time-consuming) DMRG simulation of the spin-$\frac{1}{2}$ Heisenberg tetrahedral chain with $2N$ sites ($N$ rungs) by much more effective DMRG simulation of the spin-$1$ Heisenberg chain with $N$ sites.

To accomplish this task, we have performed DMRG calculations of the spin-1 Heisenberg chain with the total number of sites $N=40$, $60$, $80$ implying the periodic boundary condition and the number of kept states up to $1200$ by making use of an open source software from Algorithms and Libraries for Physics Simulations (ALPS) project \cite{baue11}. It has been verified in Ref. \cite{hone00} that the DMRG data for the effective spin-1 Heisenberg chain of 60 sites (without any extrapolation) estimate the ground-state energy of an infinite chain with an accuracy of $10^{-4}$. From this perspective, our numerical results for the largest spin-1 Heisenberg chain of 80 sites can be straightforwardly used in order to construct the numerically exact ground-state phase diagram of the spin-$\frac{1}{2}$ Heisenberg tetrahedral chain. To eliminate finite-size effects, the magnetization curves of the spin-$\frac{1}{2}$ Heisenberg tetrahedral chain have been extrapolated to the thermodynamic limit. The rigorous numerical results obtained in this way for the spin-$\frac{1}{2}$ Heisenberg tetrahedral chain will be compared with the exact analytical results for the spin-$\frac{1}{2}$ Ising-Heisenberg tetrahedral chain in Sec. \ref{sec:comp}. 

\section{Results and discussion}
\label{sec:result}

In this section, let us proceed to a discussion of the most interesting results for the spin-$\frac{1}{2}$ Ising-Heisenberg tetrahedral chain by considering the particular case with antiferromagnetic interactions $J_x>0$, $J_z>0$ and $J_1>0$. For simplicity, our attention will be henceforth paid only to a special case of the investigated spin chain with the isotropic Heisenberg intra-rung interaction $J_x = J_z = J$, which illustrates all typical features of the more general model with the anisotropic XXZ Heisenberg interaction $J_x \neq J_z$.   

\subsection{Ground-state phase diagram}

\begin{figure}
\begin{center}
\includegraphics[width=0.5\textwidth]{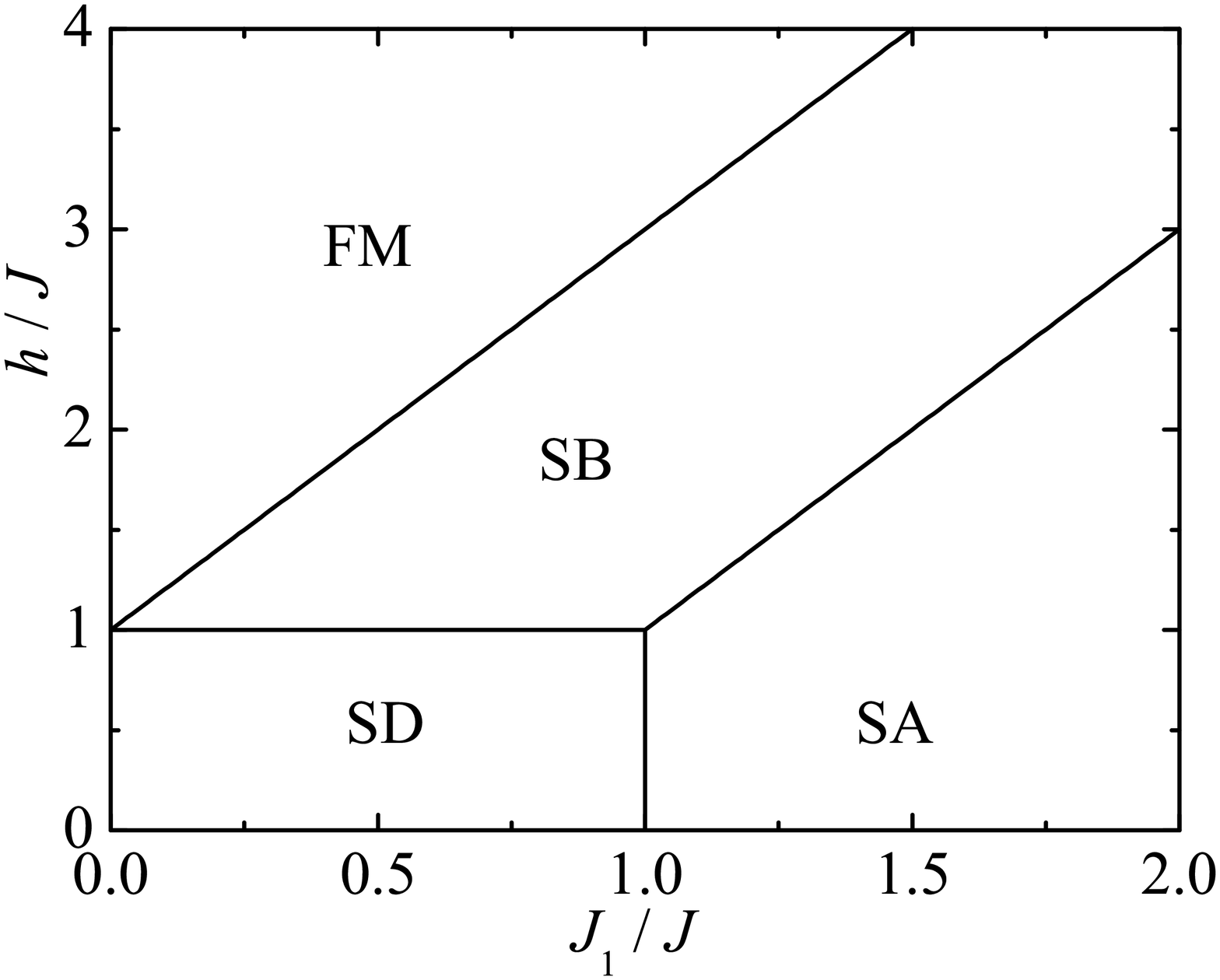}
\end{center}
\vspace{-0.6cm}
\caption{The ground-state phase diagram of the spin-$\frac{1}{2}$ Ising-Heisenberg tetrahedral chain in $J_1/J-h/J$ plane for the isotropic Heisenberg coupling 
($J_x = J_z = J$), which involves rung singlet-dimer (SD), superantiferromagnetic (SA), staggered-bond (SB), and ferromagnetic (FM) ground states.}
\label{fig2}
\end{figure} 

A diagonal form of the Hamiltonian (\ref{eq:diagham1}) can be rather straightforwardly used in order to obtain all possible ground states of the spin-$\frac{1}{2}$ Ising-Heisenberg tetrahedral chain. By inspection, one finds just four different ground states: the rung singlet-dimer (SD) state, the superantiferromagnetic (SA) state, 
the staggered-bond (SB) state and the ferromagnetic (FM) state, which are unambiguously given by the following eigenvectors
\begin{eqnarray}
|\mbox{SD}\rangle \!\!&=&\!\! \prod_{i=1}^N \frac{1}{\sqrt{2}} \left(|\! \uparrow_{1,i} \downarrow_{2,i} \rangle - |\! \downarrow_{1,i} \uparrow_{2,i} \rangle \right), 
\nonumber \\
|\mbox{SA}\rangle \!\!&=&\!\! \prod_{i=1}^{N/2} |\! \uparrow_{1,2i-1} \uparrow_{2,2i-1} \rangle \otimes |\! \downarrow_{1,2i} \downarrow_{2,2i} \rangle, \nonumber \\
|\mbox{SB}\rangle \!\!&=&\!\! \prod_{i=1}^{N/2} |\! \uparrow_{1,2i-1} \uparrow_{2,2i-1} \rangle \!\otimes\! \frac{|\! \uparrow_{1,2i} \downarrow_{2,2i} \rangle \!-\! |\! \downarrow_{1,2i} \uparrow_{2,2i} \rangle}{\sqrt{2}},  \nonumber \\
|\mbox{FM}\rangle \!\!&=&\!\! \prod_{i=1}^{N} |\! \uparrow_{1,i} \uparrow_{2,i} \rangle. 
\label{ev}
\end{eqnarray}
The ground-state phase diagram involving all available ground states is depicted in Fig. \ref{fig2}. The unique SD state with all Heisenberg bonds (rungs) in the singlet-dimer state becomes the ground state in a parameter space delimited by the conditions $J_1<J$ and $h<J$. If the conditions $J_1>J$ and $h<2J_1-J$ are met, the ground state is formed by the two-fold degenerate SA spin arrangement with a regular alternation the fully polarized Heisenberg rungs into two opposite directions. Evidently, the total magnetization of SD and SA ground states equals zero. As long as the reverse conditions $J_1<J$, $h>J$ or $J_1>J$, $h>2J_1-J$ are fulfilled, the two-fold degenerate SB state with a regular alternation of the singlet dimers and polarized triplets occurs as a result of the field-induced transition from the SD or SA ground state. According to this, the spin-$\frac{1}{2}$ Ising-Heisenberg tetrahedral chain must exhibit in a zero-temperature magnetization curve an intermediate plateau at half of the saturation magnetization due to the symmetry-broken magnetic structure of the SB ground state in which half of the Heisenberg rungs is polarized by the external magnetic field. Finally, the model under investigation shows a transition towards the fully polarized FM ground state as soon as the saturation field $h = 2 J_1 + J$ is reached. 

Note furthermore that the macroscopic degeneracy at the boundaries between different phases can be found at the relevant field-induced transitions. At the SA-SB boundary each second rung can stay in two different states (singlet and fully polarized) leading to the macroscopic degeneracy of $2^{N/2}$. The degenerate states at the SD-SB and SB-FM boundaries can be described by the effective one-dimensional hard-dimer model using the procedure quite analogous to that one reported in Ref. \cite{derz07}. As a result, the macroscopic degeneracy is $((1+\sqrt{5})/2)^{N}$ for this case. The obtained macroscopic degeneracy leads to a residual ground-state entropy, which will be more thoroughly examined in Sec.~\ref{adiabatic}.

\subsection{Magnetization process}

\begin{figure}
\begin{center}
\includegraphics[width=0.49\textwidth]{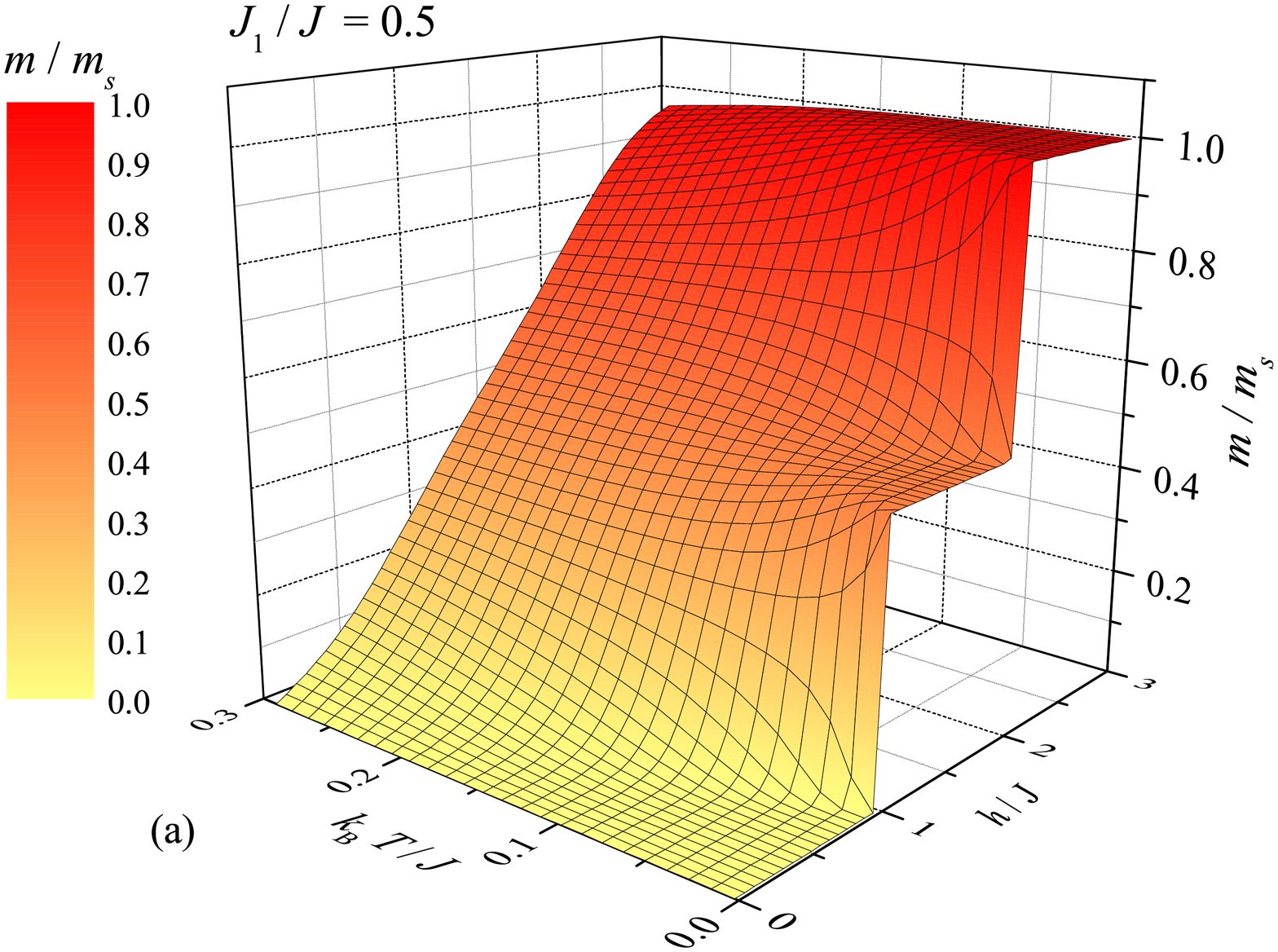}
\includegraphics[width=0.49\textwidth]{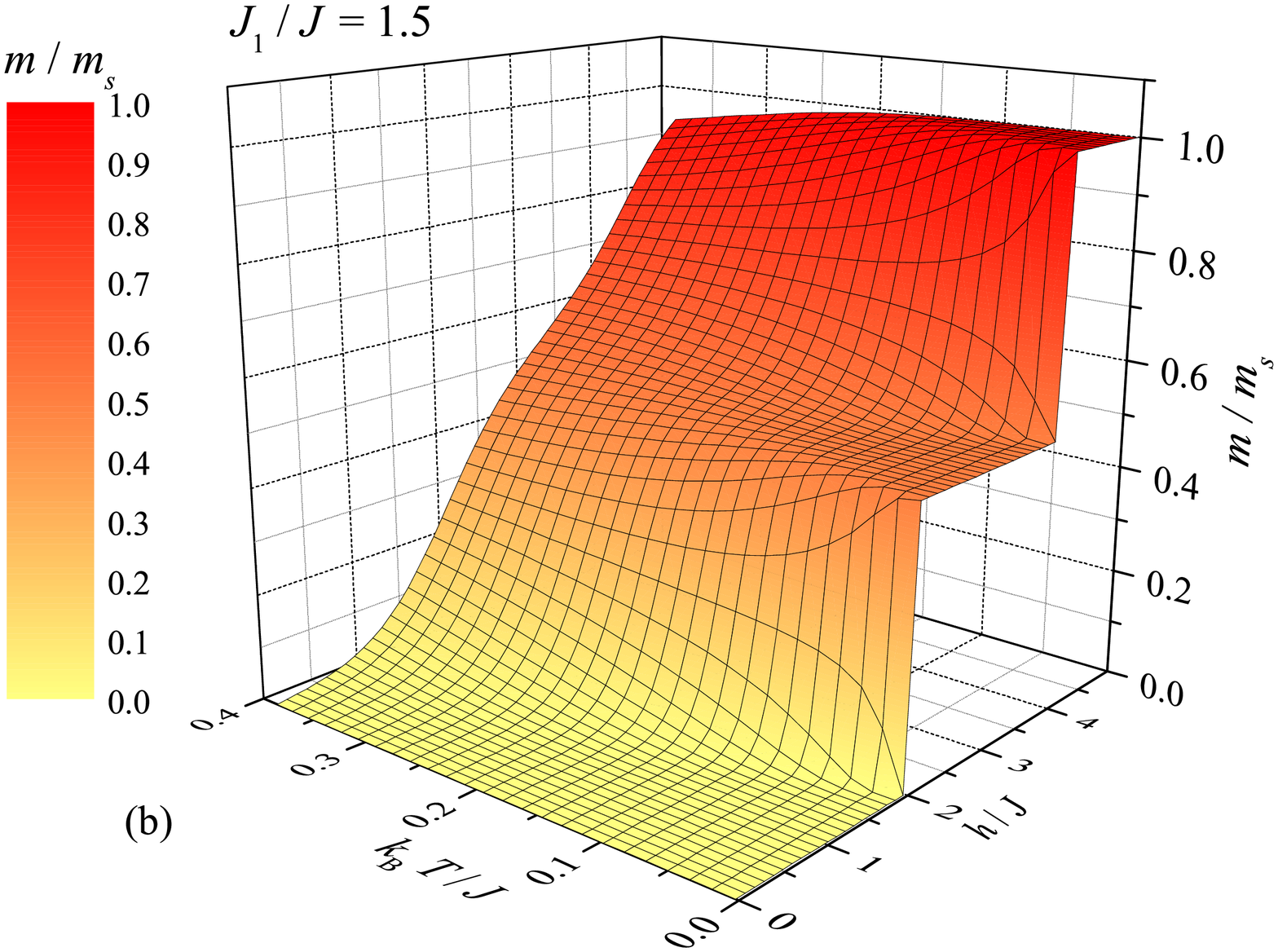}
\end{center}
\vspace{-0.6cm}
\caption{3D plot of the magnetization normalized with respect to the saturation value against temperature and magnetic field for the isotropic Heisenberg coupling ($J_x = J_z = J$) and two different values of the interaction ratio: (a) $J_1/J = 0.5$; (b) $J_1/J = 1.5$.}
\label{fig3}
\end{figure}

Let us proceed to a detailed examination of the magnetization process to verify the presence of the intermediate plateau at one-half of the saturation magnetization, which would correspond to a field-induced stabilization of the SB ground state. Fig. \ref{fig3} shows a three-dimensional (3D) plot of the magnetization as a function of the dimensionless magnetic field and temperature for two alternative choices of the interaction ratio $\frac{J_1}{J} = 0.5$ and $1.5$ driving the investigated spin chain in the zero-field limit towards the SD and SA ground state, respectively. As one can see, the low-temperature magnetization curves clearly serve in evidence of the one-half magnetization plateau irrespective of whether the SD or SA state is being the respective zero-field ground state. The stepwise magnetization curve with a steep increase of the magnetization observable in a vicinity of the relevant transition fields is generally smoothened by increasing temperature, whereas the thermally invoked smoothening occurs faster for the unique quantum SD state than the classical but doubly degenerate SA state. 

\subsection{Bipartite entanglement}

\begin{figure}
\begin{center}
\includegraphics[width=0.49\textwidth]{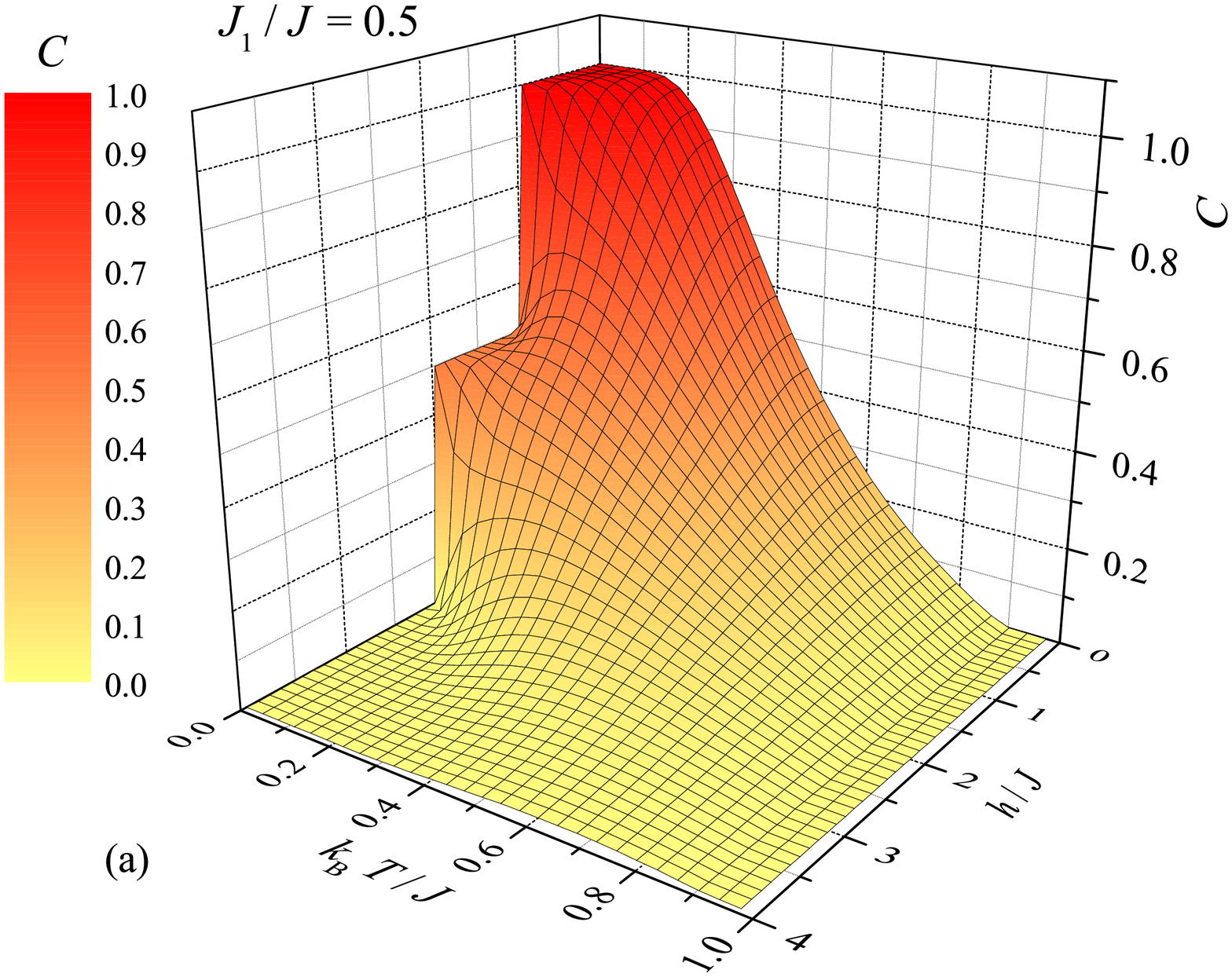}
\includegraphics[width=0.49\textwidth]{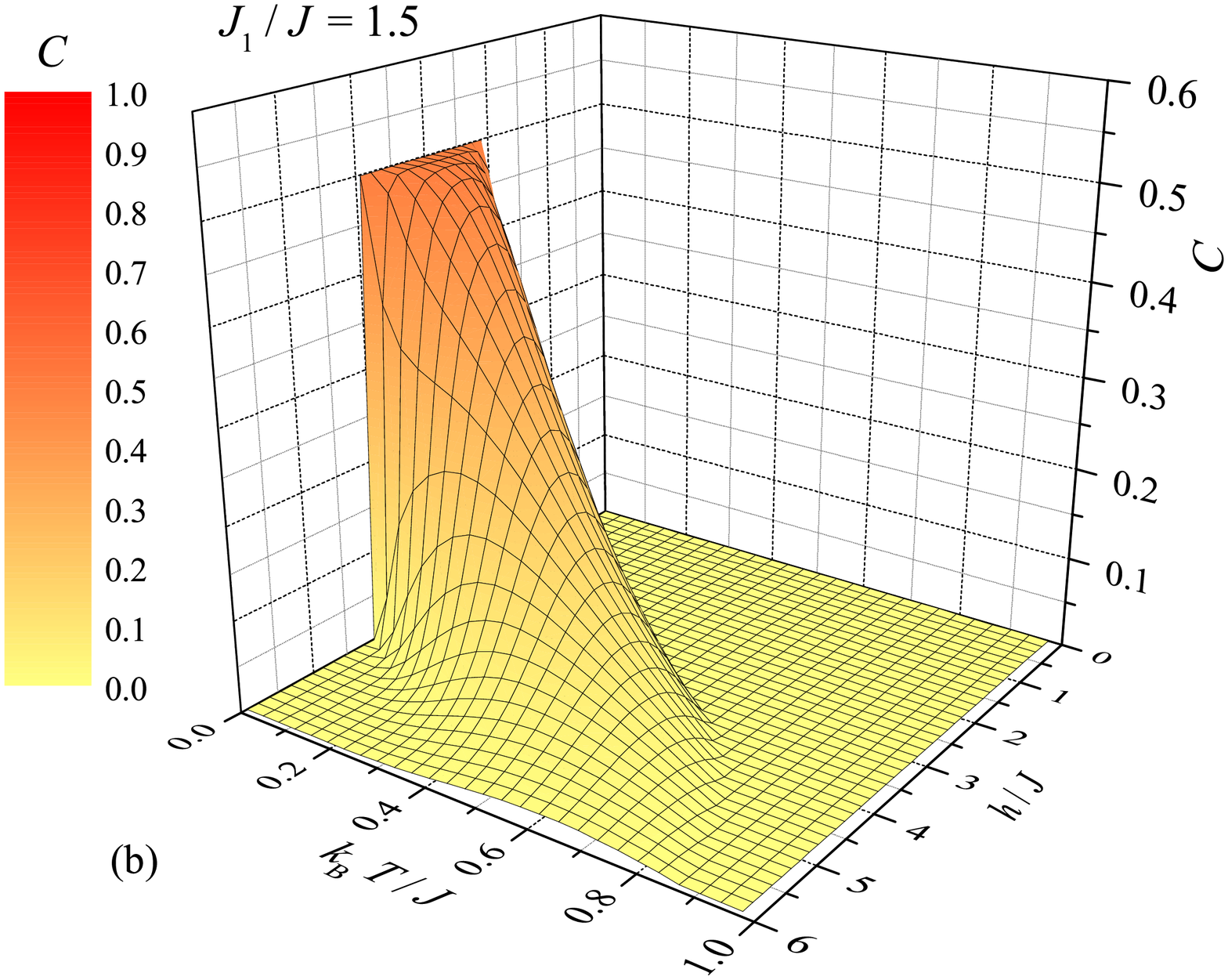}
\end{center}
\vspace{-0.6cm}
\caption{3D plot of the concurrence against temperature and magnetic field for the isotropic Heisenberg coupling ($J_x = J_z = J$) and two different values of the interaction ratio: (a) $J_1/J = 0.5$; (b) $J_1/J = 1.5$.}
\label{fig4}
\end{figure}

Although two magnetization curves displayed in Fig. \ref{fig3} are qualitatively the same regardless of whether the quantum SD or classical SA spin arrangement is realized at low enough magnetic fields, one may take advantage of the concurrence for two spins coupled by the Heisenberg interaction to discern both different magnetization scenarios. To bring an insight into the degree of quantum correlations during the magnetization process, the analogous 3D plot of the concurrence against temperature and magnetic field is depicted in Fig. \ref{fig4} as formerly presented for the magnetization. Evidently, the former particular case shown in Fig. \ref{fig4}(a) is consistent with the existence of the maximally entangled SD ground state in a low-field region, while the latter particular case depicted in Fig. \ref{fig4}(b) demonstrates a lack of quantum correlations at sufficiently low fields due to the classical SA ground state. If the SD state forms the zero-field ground state, the magnetic field generally acts in conjunction with temperature to diminish the concurrence (Fig. \ref{fig4}(a)). There is however one fundamental difference between temperature and magnetic-field effect upon the quantum entanglement. While the field-induced changes of the concurrence are abrupt, the concurrence diminishes more continuously owing to the increasing temperature  until it completely disappears at some threshold temperature. The macroscopic magnetization jumps and plateaux are accordingly reflected in the respective entanglement jumps and plateaux as well \cite{bose02}. Another interesting observation can be made for the other particular case with the classical SA state as the zero-field ground state (Fig. \ref{fig4}(b)). Under this condition, the rising magnetic field causes a peculiar abrupt increase of the concurrence on behalf of the field-induced transition towards the SB ground state. It is noteworthy that the concurrence achieves exactly half of its maximum value within the SB ground state, because the maximally entangled singlet dimers regularly alternate with the fully polarized (disentangled) triplet states \cite{bose02}. The fact that the external field is responsible for an appearance of quantum correlations above the classical  SA ground state represents a quite unexpected finding, since the longitudinal magnetic field usually destroys quantum correlations.

\subsection{Adiabatic demagnetization}
\label{adiabatic}

\begin{figure}
\begin{center}
\includegraphics[width=0.45\textwidth]{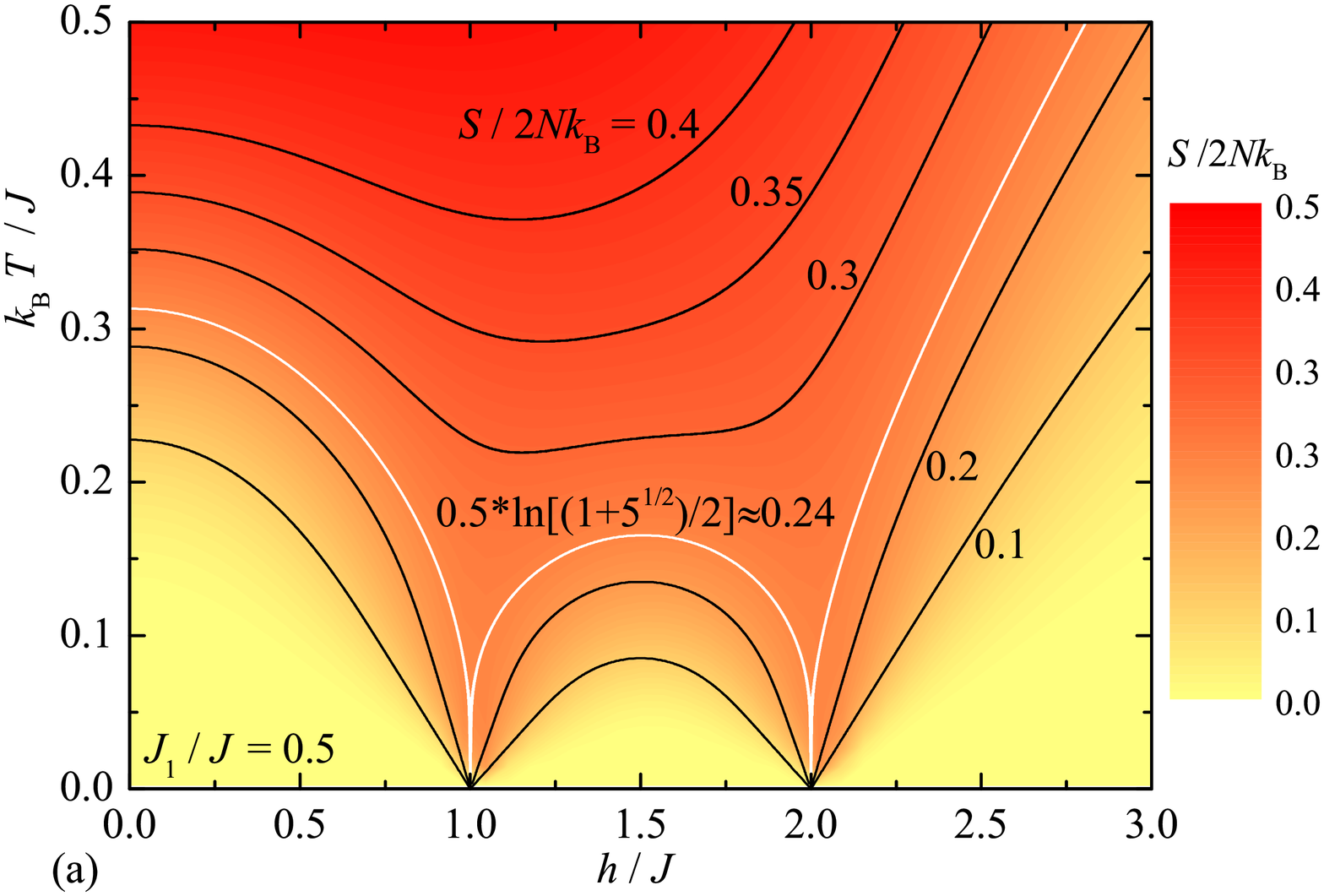}
\includegraphics[width=0.45\textwidth]{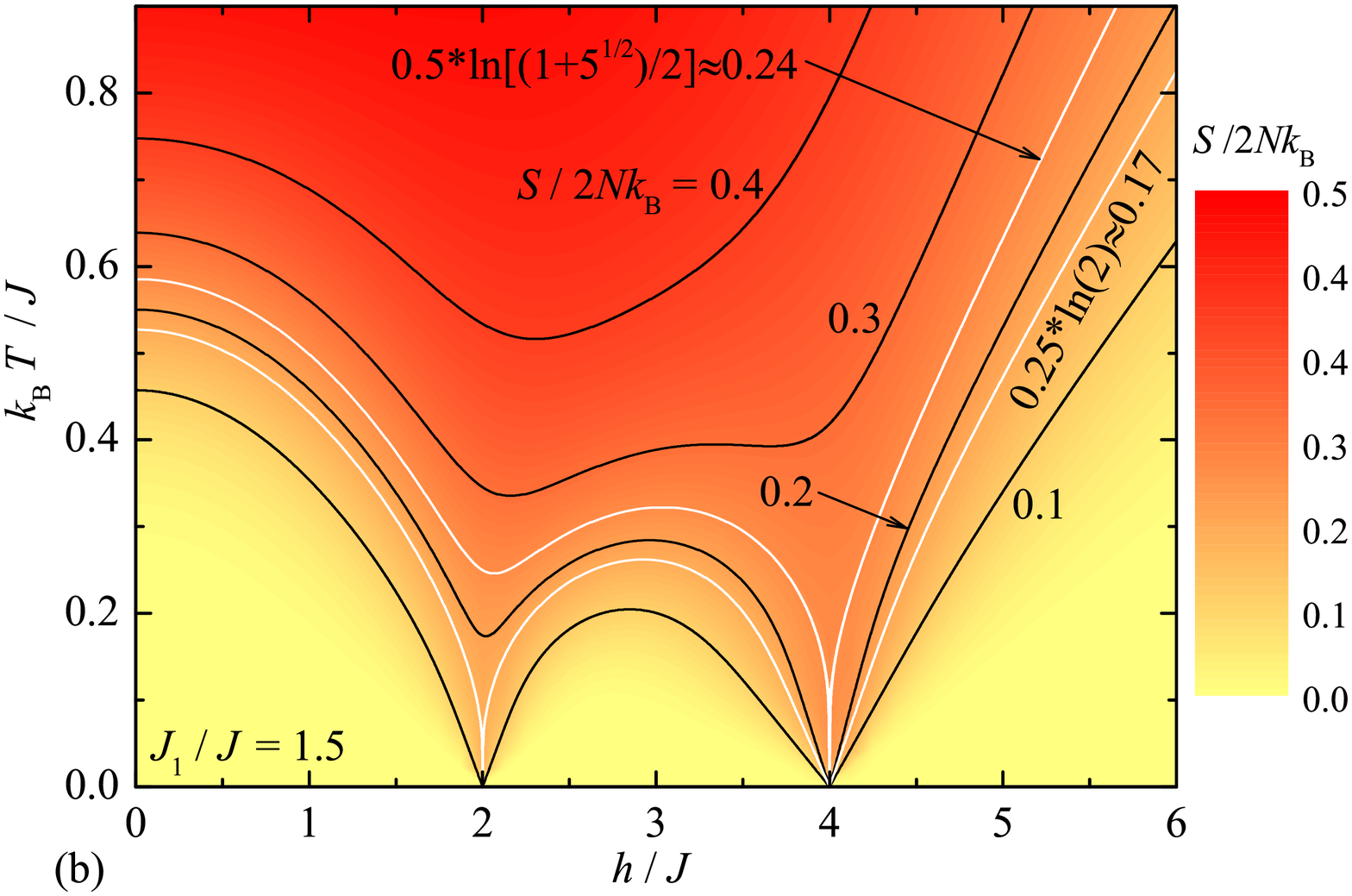}
\end{center}
\vspace{-0.6cm}
\caption{Temperature as a function of the magnetic field by keeping entropy constant and selecting two different values of the interaction ratio: (a) $J_1/J = 0.5$; (b) $J_1/J = 1.5$. White lines correspond to the particular cases with the largest magnetocaloric effect.}
\label{fig5}
\end{figure}

Recently, it has been demonstrated that several frustrated spin systems may exhibit an enhanced magnetocaloric effect during the adiabatic demagnetization, which might be of practical importance for low-temperature magnetic refrigeration \cite{zhit03,zhit04,hone06,schn07,rich05,derz06}. Owing to this fact, let us also investigate the adiabatic demagnetization of the spin-$\frac{1}{2}$ Ising-Heisenberg tetrahedral chain under the adiabatic (isentropic) conditions. Fig. \ref{fig5} illustrates typical isentropic changes of temperature upon varying the external magnetic field for the two different magnetization scenarios discussed previously. A sequence of two field-induced transitions FM$\to$SB$\to$SD upon decreasing the magnetic field is reflected in the relevant temperature changes shown in Fig. \ref{fig5}(a). It is quite obvious from this figure that the most prominent magnetocaloric effect can be detected just if the entropy is selected sufficiently close to the value $S = N k_{\rm B} \ln (\frac{1 + \sqrt{5}}{2})$, under which temperature vanishes infinitely fast as the external field approaches one of the two transition fields. The other particular case shown in Fig. \ref{fig5}(b) reflects another possible sequence of two field-induced transitions FM$\to$SB$\to$SA, which displays an enhanced magnetocaloric effect provided that the entropy is kept either close to the value $S = N k_{\rm B} \ln (\frac{1 + \sqrt{5}}{2})$ or $\frac{1}{2} \ln (2)$. The former value $S = N k_{\rm B} \ln (\frac{1 + \sqrt{5}}{2})$ ensures an efficient cooling in a vicinity of the FM$\to$SB transition, while the latter value $S = N k_{\rm B} \frac{1}{2} \ln (2)$ affords an effective cooling nearby the SB$\to$SA transition. The optimal entropy values for an effective cooling are consistent with the respective ground-state degeneracies reported previously for the relevant field-induced transitions.

\subsection{Specific heat}

\begin{figure}
\begin{center}
\includegraphics[width=0.49\textwidth]{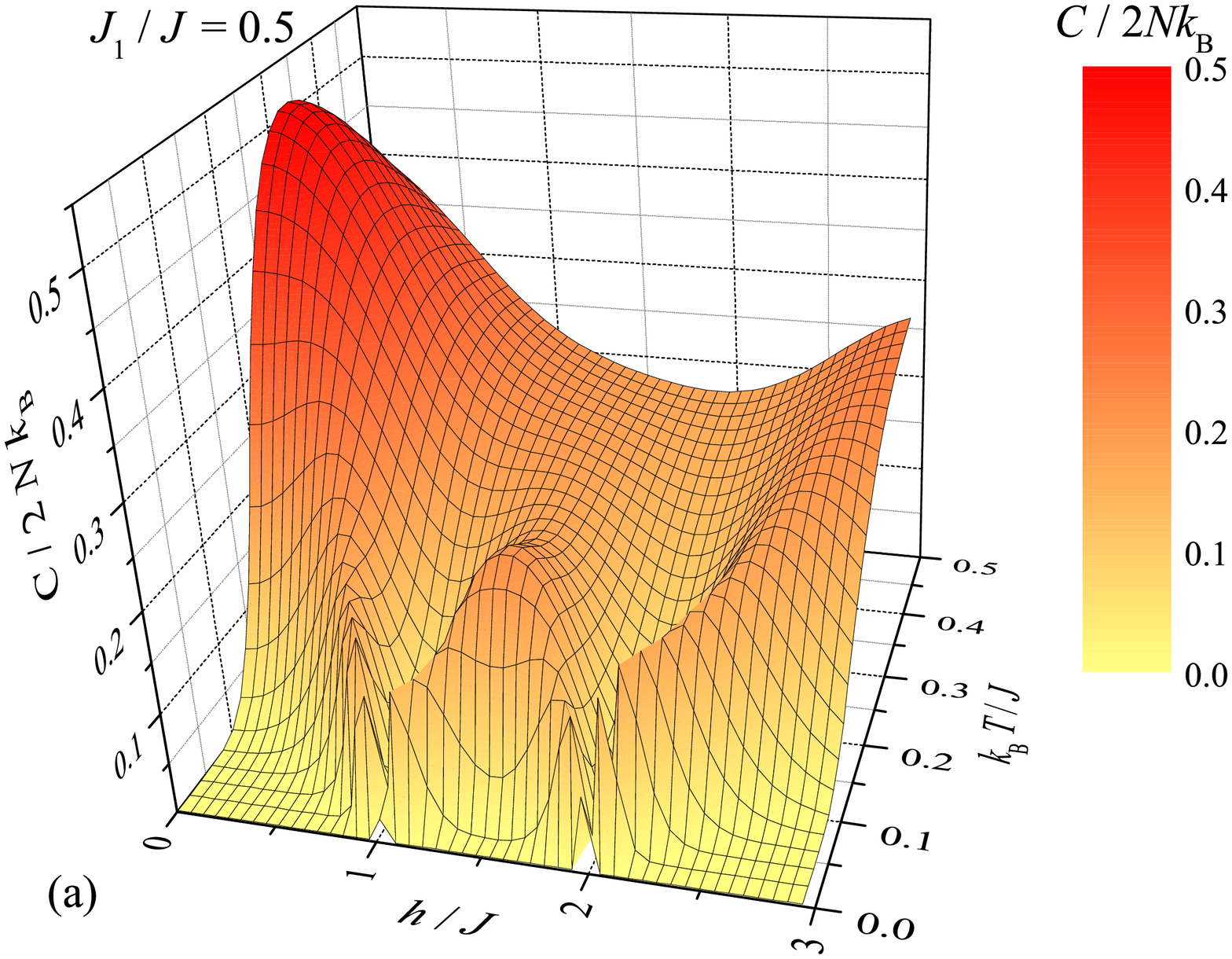}
\includegraphics[width=0.49\textwidth]{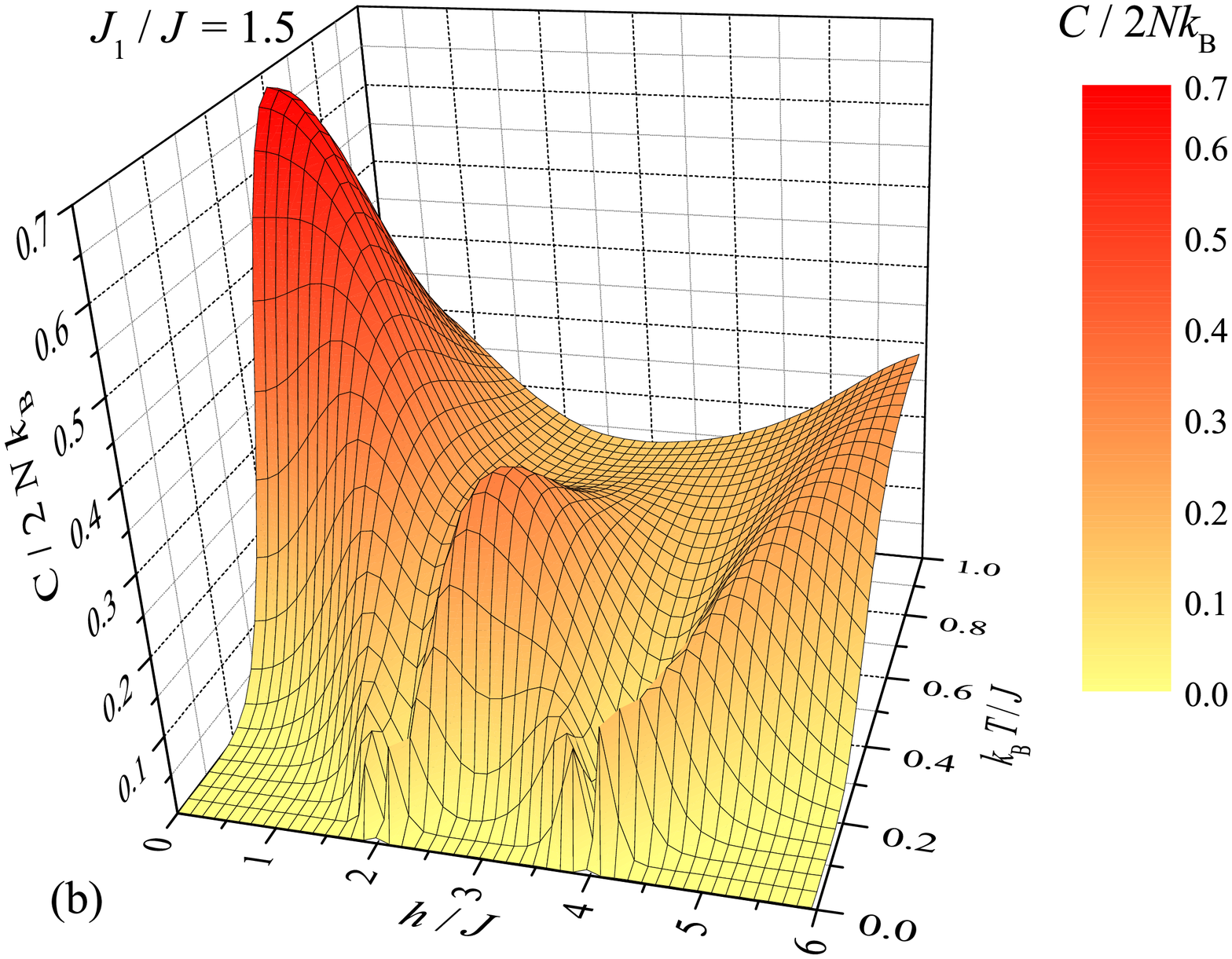}
\end{center}
\vspace{-0.6cm}
\caption{3D plot of the specific heat against temperature and magnetic field for the isotropic Heisenberg coupling ($J_x = J_z = J$) and two different values of the interaction ratio: (a) $J_1/J = 0.5$; (b) $J_1/J = 1.5$.}
\label{fig6}
\end{figure}

Last but not least, let us turn our attention to typical dependences of the specific heat on temperature and magnetic field as displayed in Fig. \ref{fig6}. Thermal variations of zero-field specific heat exhibit a single rounded Schottky-type maximum, which is slightly higher for the latter particular case with the SA ground state (Fig. \ref{fig6}(b)) in comparison with the former particular case with the SD ground state (Fig. \ref{fig6}(a)). As far as the low-temperature field dependence of the specific heat is concerned, the specific heat displays as a function of the external magnetic field two marked peaks around the transition fields towards the intermediate SB state. In the consequence of that, the simple thermal dependence of the zero-field specific heat passes to a more complex temperature dependence with two more or less separated round maxima by turning on the external magnetic field, whereas the most pronounced double-peak specific heat curves can be found in a neighborhood of the transition fields.  

\section{Ising-Heisenberg vs. Heisenberg tetrahedral chain}
\label{sec:comp}

In this section, the ground-state phase diagram and zero-temperature magnetization curves of the spin-$\frac{1}{2}$ Ising-Heisenberg tetrahedral chain will be confronted with the analogous results obtained for the fully quantum spin-$\frac{1}{2}$ Heisenberg tetrahedral chain with the help of DMRG method. Recall that the DMRG calculations were performed for the effective spin-1 Heisenberg chain with up to 80 sites and 1200 kept states, which afford sufficiently precise estimate of the ground-state energy to construct the numerically exact ground-state phase diagram of the spin-$\frac{1}{2}$ Heisenberg tetrahedral chain \cite{hone00}. 

\begin{figure}
\begin{center}
\includegraphics[width=0.49\textwidth]{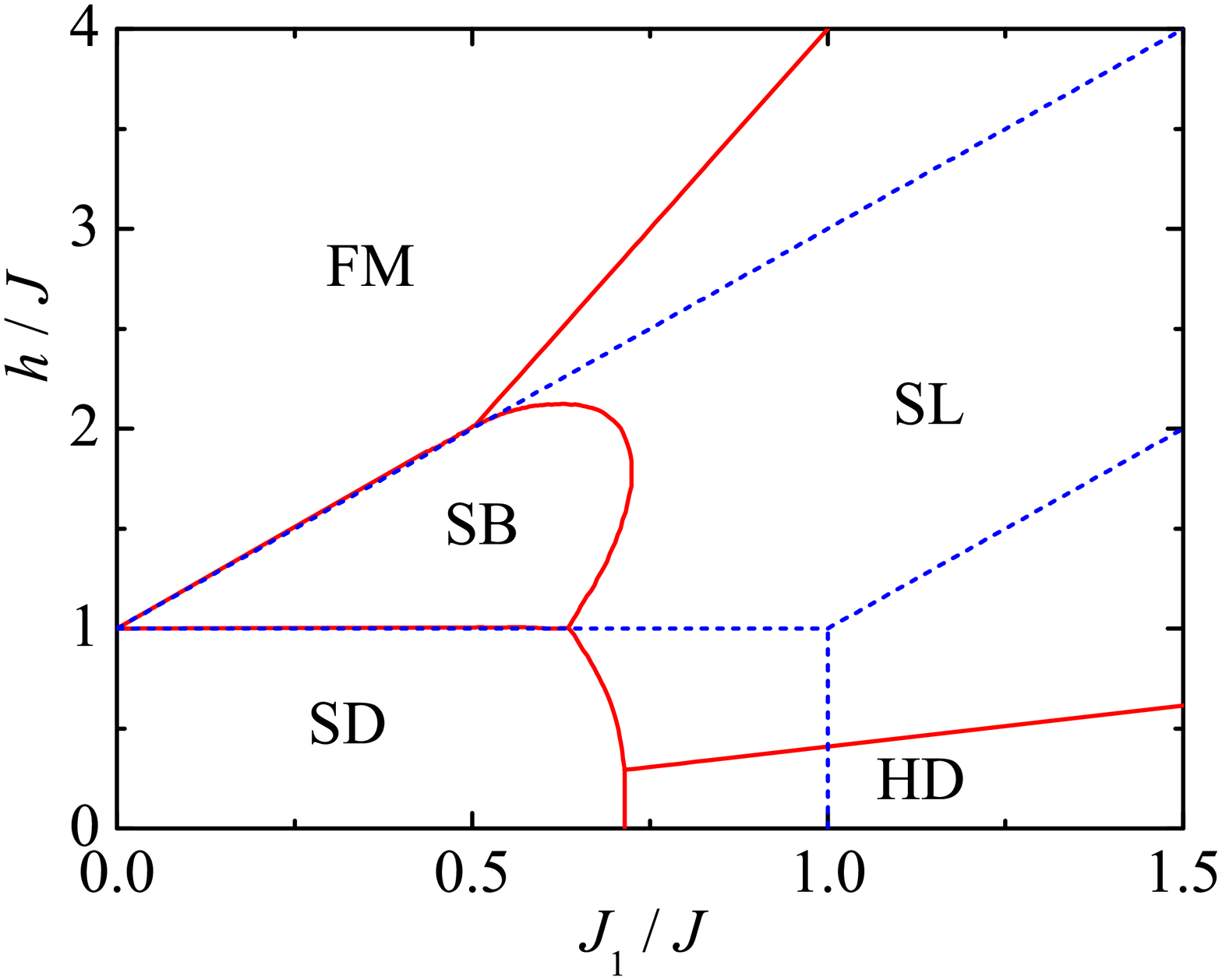}
\end{center}
\vspace{-0.6cm}
\caption{The ground-state phase diagram of the Heisenberg tetrahedral chain (red solid lines) is plotted along with the ground-state phase boundaries of the Ising-Heisenberg tetrahedral chain (blue broken lines). For better clarity, the notation for individual ground states is given just for the Heisenberg tetrahedral chain.}
\label{fig7}
\end{figure}

The ground-state phase diagrams of the spin-$\frac{1}{2}$ Ising-Heisenberg and Heisenberg tetrahedral chain are plotted together in Fig. \ref{fig7}. It is quite obvious from this figure that both ground-state phase diagrams exactly coincide for sufficiently weak values of the interaction ratio $\frac{J_1}{J} \leq 0.5$. Under this condition, the zero-temperature magnetization curve of both investigated spin-chain models exhibits strict magnetization plateaux at zero and half of the saturation magnetization, which end up at macroscopic magnetization jumps associated with a closing singlet-triplet gap and respectively, the breakdown of the state with the highest density of independent localized one-magnon states \cite{rich05,derz06,derz07,derz10,magn}. The high numerical precision of the constructed ground-state phase diagram is also confirmed by rigorous analytical results for two associated transition fields $h_{c1} = J$ and $h_{c2} = J + 2J_1$ of the spin-$\frac{1}{2}$ Heisenberg tetrahedral chain, which has been proved by the strong-coupling approach in the parameter region $\frac{J_1}{J} \leq 0.5$ \cite{mila98}. 

\begin{figure*}
\begin{center}
\includegraphics[width=0.53\textwidth]{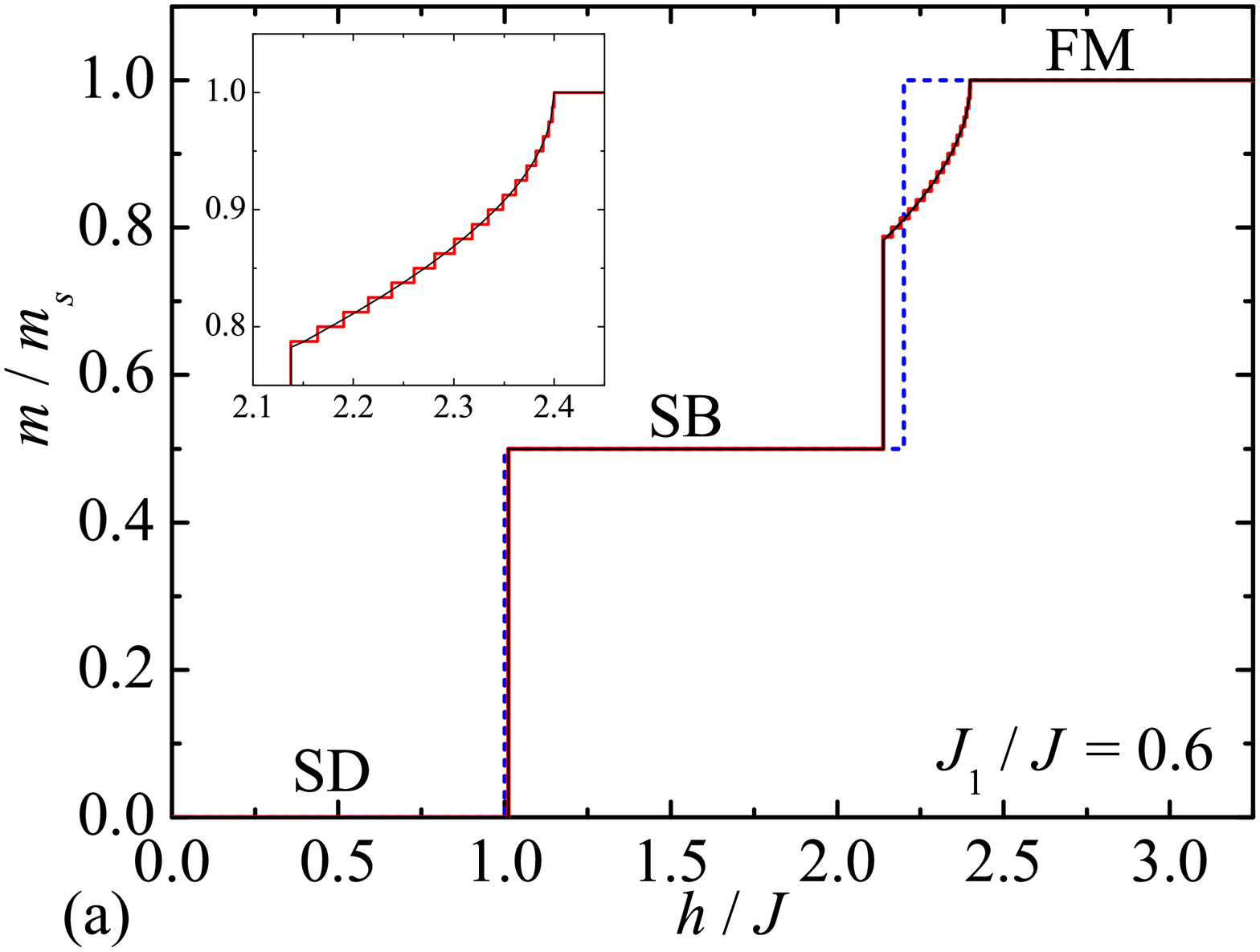}
\hspace*{-1.5cm}
\includegraphics[width=0.53\textwidth]{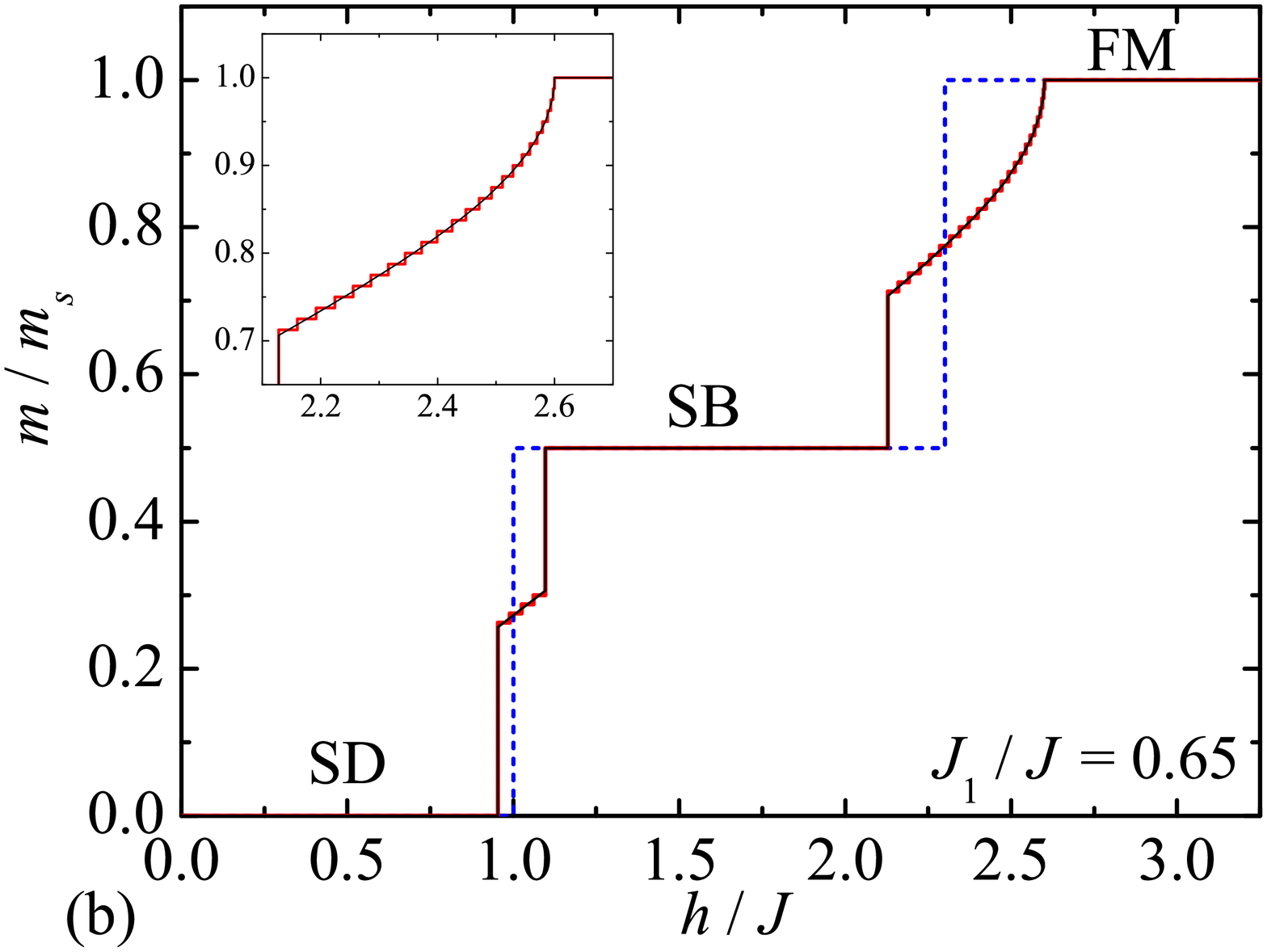}
\includegraphics[width=0.53\textwidth]{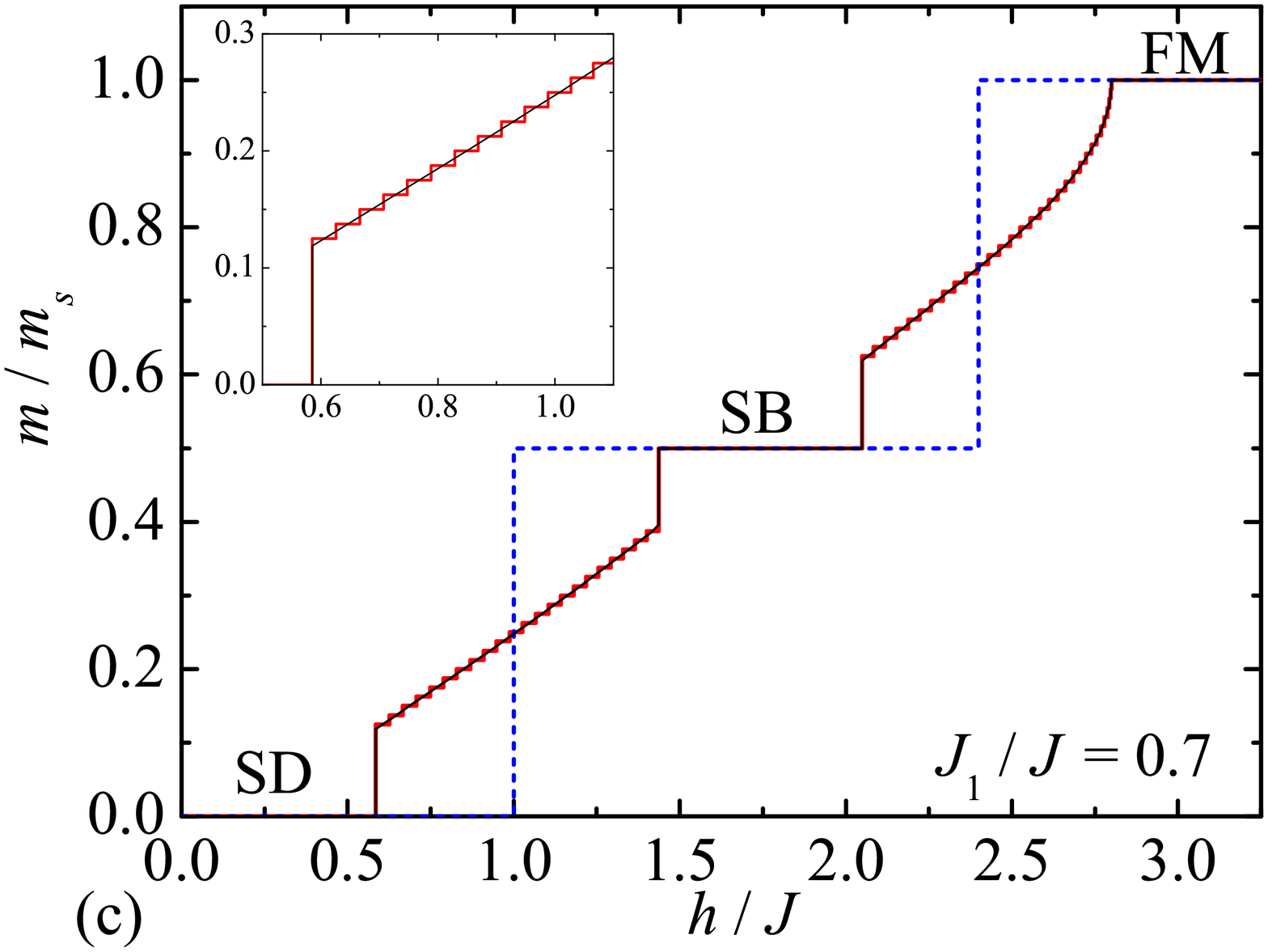}
\hspace*{-1.5cm}
\includegraphics[width=0.53\textwidth]{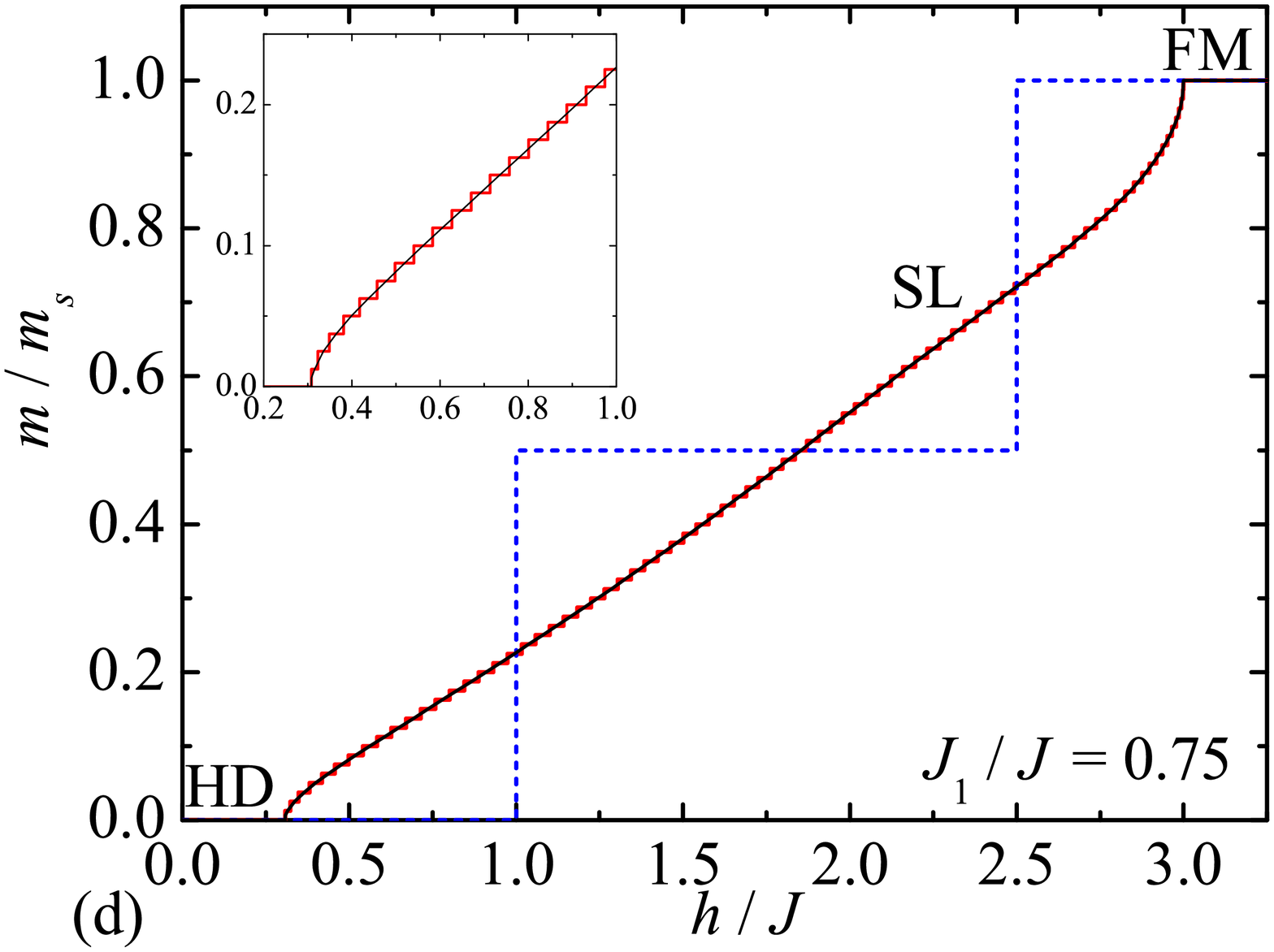}
\end{center}
\vspace{-0.6cm}
\caption{Zero-temperature magnetization curves of the Ising-Heisenberg tetrahedral chain (blue broken lines) are confronted with the magnetization curves of the full quantum Heisenberg tetrahedral chain (solid lines). The thick stepwise (red) line shows the magnetization curves of the spin-$\frac{1}{2}$ Heisenberg tetrahedral chain of 160 sites (80 rungs) obtained from the DMRG data of the effective spin-1 Heisenberg chain of 80 sites, while the smooth thin (black) line shows the magnetization curve extrapolated for the infinite system. Insets show both these largely overlapping  curves in an enlargened scale. The magnetization curves are presented for four different values of the interaction ratio: (a) $J_1/J = 0.6$; (b) $J_1/J = 0.65$; (c) $J_1/J = 0.7$; (d) $J_1/J = 0.75$.}
\label{fig8}
\end{figure*}

The first difference between the ground-state phase diagrams of the spin-$\frac{1}{2}$ Ising-Heisenberg and Heisenberg tetrahedral chain can be found by considering  $\frac{J_1}{J} \in (0.50, 0.64)$ when the latter Heisenberg chain exhibits a gapless Luttinger spin-liquid (SL) ground state just before reaching the saturation. The respective zero-temperature magnetization curve supporting this statement is presented in Fig. \ref{fig8}(a). If the interaction ratio is selected from the interval $\frac{J_1}{J} \in (0.64, 0.71)$, the gapless Luttinger SL state also emerges in between SD and SB ground states instead of a direct field-induced transition in between those ground states. However, the magnetization plateaux at zero and one-half of the saturation magnetization still persist in a magnetization process of the Ising-Heisenberg and Heisenberg tetrahedral chains including their SD and SB nature. The relevant zero-temperature magnetization curves, which illustrate a continuous change of the magnetization in a parameter region in between zero and one-half magnetization plateaux are depicted in Fig. \ref{fig8}(b)-(c). Finally, the most crucial difference between a magnetic behavior of both considered spin-chain models can be detected for $\frac{J_1}{J} \geq 0.71$. In this parameter space, the respective ground state of the spin-$\frac{1}{2}$ Heisenberg tetrahedral chain is formed by the Haldane-like (HD) state with the energy gap $\Delta = 0.41 J_1$ unlike the other gapped SD and SA ground states of the spin-$\frac{1}{2}$ Ising-Heisenberg tetrahedral chain. Consequently, the magnetization of the spin-$\frac{1}{2}$ Heisenberg tetrahedral chain remains zero until the magnetic field closes the Haldane gap and then it continuously rises with the external field until the saturation magnetization is reached (see Fig. \ref{fig8}(d)).  

\section{Conclusion}
\label{sec:conc}

In summary, we have provided a detailed study of the spin-$\frac{1}{2}$ Ising-Heisenberg tetrahedral chain, which has been exactly solved by taking advantage of local conservation for the total spin on the Heisenberg bonds and transfer-matrix technique. In particular, we have examined the ground state, magnetization process, quantum entanglement, magnetocaloric effect and overall thermodynamics of this exactly solvable spin-chain model. Exact results for the ground state and zero-temperature magnetization curves of the spin-$\frac{1}{2}$ Ising-Heisenberg tetrahedral chain have been also confronted with the corresponding results of the analogous but fully quantum spin-$\frac{1}{2}$ Heisenberg tetrahedral chain obtained within the framework of DMRG calculations. 

The most interesting finding stemming from the present study can be viewed in providing correlation between the magnetic behavior and thermal entanglement of the spin-$\frac{1}{2}$ Ising-Heisenberg tetrahedral chain. It has been demonstrated that the magnetization plateaux observable at sufficiently low temperatures manifest themselves in respective plateaux of the concurrence, which has been exploited for measuring the bipartite entanglement between two spins coupled by the Heisenberg interaction. The exactly solved Ising-Heisenberg tetrahedral chain thus verifies the well-known dictum that the ground-state concurrence and other entanglement witnesses undergo a discontinuous change across the first-order quantum phase transition driven by the external magnetic field \cite{amic08,trib09,bose05}. In addition, the abrupt temperature-induced changes of the concurrence in a close vicinity of the critical fields afford a clear signature of first-order quantum phase transtions at finite temperatures. The spin-$\frac{1}{2}$ Ising-Heisenberg tetrahedral chain exhibits a lot of other interesting features including the enhanced magnetocaloric effect or double-peak specific heat curves. It should be stressed, moreover, that the exact results for thermodynamic quantities of the spin-$\frac{1}{2}$ Ising-Heisenberg tetrahedral chain might also be useful for a better understanding of the magnetic behavior of the quantum spin-$\frac{1}{2}$ Heisenberg tetrahedral chain at finite temperatures on assumption that the intra-rung coupling is at least twice as large as the inter-rung coupling.

\begin{acknowledgments}
J.S. acknowledges the financial support provided by the grant of The Ministry of Education, Science, Research and Sport of the Slovak Republic under the contract No. VEGA 1/0234/12 and by the grants of the Slovak Research and Development Agency under the contract Nos. APVV-0132-11 and APVV-0097-12. O.R. thanks the Brazilian agencies CNPq and FAPEMIG. M.L.L. also acknowledges the financial support of CNPq, CAPES (Brazilian research agencies) and FAPEAL (Alagoas state research agency). 
\end{acknowledgments}


\begin{thebibliography}{100}
\bibitem{magn}
The SB state actually represents the state with the highest density of independent localized one-magnon states, 
which can be trapped on the Heisenberg rungs on a ferromagnetic background \cite{rich05,derz06,derz07,derz10}.
\end{thebibliography}
\end{document}